\documentclass[conference]{IEEEtran}
\IEEEoverridecommandlockouts
\usepackage{cite}
\usepackage{amsmath,amssymb,amsfonts,pifont}
\usepackage{algorithmic}
\usepackage{graphicx}
\usepackage{color}
\usepackage{textcomp}
\usepackage{multirow}
\usepackage{comment}
\usepackage{xcolor}
\usepackage[hang,small,bf]{caption}
\usepackage[subrefformat=parens]{subcaption}
\newcommand{\cmark}{\ding{51}}%
\newcommand{\xmark}{\ding{55}}%

\newcommand{\add}[1]{\textcolor{black}{#1}}	
\newcommand{\erase}[1]{\if0{#1}\fi}	

\def\BibTeX{{\rm B\kern-.05em{\sc i\kern-.025em b}\kern-.08em
    T\kern-.1667em\lower.7ex\hbox{E}\kern-.125emX}}
\begin{document}

\title{ZEL: Net-Zero-Energy Lifelogging System using Heterogeneous Energy Harvesters}

\author{
\IEEEauthorblockN{
Anonymized authors
}
\IEEEauthorblockA{
}
}

\author{
\IEEEauthorblockN{
Mitsuru Arita$^{1}$, Yugo Nakamura$^{1,2}$, Shigemi Ishida $^{3}$, Yutaka Arakawa$^{1}$\\
}
\IEEEauthorblockA{
$^1$Kyushu University, Fukuoka 819-0395, Japan\\
$^2$JST Presto,
Chiyoda-ku, Tokyo 102-0076, Japan\\
$^3$Future University Hakodate, Hokkaido 041-8655, Japan\\
Email: {arita.mitsuru, yugo, yutaka, ishida}@arakawa-lab.com}
}


\maketitle

\begin{abstract}

We present ZEL, the first net-zero-energy lifelogging system that allows office workers to collect semi-permanent records of when, where, and what activities they perform on company premises.
ZEL achieves high accuracy lifelogging by using heterogeneous energy harvesters with different characteristics.
The system is based on a 192-gram nametag-shaped wearable device worn by each employee that is equipped with two \erase{voltage detectors} \add{comparators} to enable seamless switching between system states, thereby minimizing the battery usage and enabling net-zero-energy, semi-permanent data collection.
To demonstrate the effectiveness of our system, we conducted data collection experiments with 11 participants in a practical environment and found that the person-dependent (PD) model achieves an 8-place recognition accuracy level of 87.2\% (weighted F-measure) and a static/dynamic activities recognition accuracy level of 93.1\% (weighted F-measure).
Additional testing confirmed the practical long-term operability of the system and showed it could achieve a zero-energy operation rate of 99.6\% i.e., net-zero-energy operation.

\end{abstract}

\begin{IEEEkeywords}
Energy harvesting, Net-zero-energy system, 
\erase{Wearable, Solar, Lifelog, Place recognition, Activity recognition}
\add{Lifelog, Place recognition, Activity recognition}
\end{IEEEkeywords}

\section{introduction} \label{sec:intro}
\erase{The number of patients suffering from lifestyle-related diseases is increasing and has become a global social problem.
Office workers, in particular, tend to suffer from elevated stress and a lack of exercise, and numerous studies conducted on lifestyles and health show they are at risk of suffering from lifestyle-related diseases.}
Reviewing lifestyle behaviors using lifelog data is one way to help prevent \erase{them} \add{lifestyle-related diseases}.
For example, the amount of daily exercise or the number of times a person visits a smoking area can be objectively monitored, potentially leading that person to take the first steps towards improving his or her lifestyle.
However, to be effective, a lifelog requires three types of information: when, where, and what activity.

\erase{
Global Positioning System (GPS) data are often used for place recognition.
However, while this use can achieve high accuracy outdoors, it is less effective indoors because of the influence of obstructions.
For office workers, who spend most of their day in a single building, when it is important to know where they are indoors, positioning methods using beacons with Wi-Fi, Bluetooth Low Energy (BLE), etc.\ have been studied\cite{zou2017winips,robesaat2017improved,flab/ishida16:iiai_aai_eskm,flab/ishida18:iiai_aai_eskm}.
However, all current radio-wave-based methods require exclusive beacons and wireless communication devices such as smartphones to send and receive radio wave signals.
This leads to problems with battery consumption and beacon installation costs. 
On the other hand, 
accelerometers are often used in activity recognition, and it is well known that attaching an accelerometer to a person makes it possible to recognize human activities such as resting, walking, or traveling upstairs/downstairs \cite{bao2004activity, chen2012sensor, sztyler2017position, nakamura2019waistonbelt}.
}

\add{
Global Positioning System (GPS) data and accelerometers are often used for lifelogging.
These sensors can sense fine positioning and human activity such as resting, walking, or traveling upstairs/downstairs \cite{bao2004activity, chen2012sensor, sztyler2017position, nakamura2019waistonbelt}.
}

In the case of lifelogging, high-resolution positioning or activity recognition is not important, but rough log such as classrooms or restrooms.
Therefore, the above high powered sensors not only shorten the battery lifetime, but are also over-performance for lifelogging.

\erase{
However, a sufficient sampling rate and high power levels are required.
Therefore, it is difficult to achieve zero-energy lifelogging using inertial activity sensors that require high power consumption. However, in the case of lifelogging, high-resolution positioning is not as important as it is in the case of beacon-based methods, and a history of time spent in places such as classrooms or restrooms is sufficient.
}

\begin{table*}[bt]
\centering
\caption{Summary of work related to using energy harvesters as context recognition sensors and our proposed ZEL system.}
\begin{tabular}{lllcccc} \hline
\multirow{2}{*}{year} & \multirow{2}{*}{References} &
\multirow{2}{*}{Energy harvesters} & 
\multicolumn{2}{c}{Recognition target} &
\multicolumn{1}{l}{\multirow{2}{*}{\begin{tabular}[c]{@{}l@{}}Net-zero-energy\\Design\end{tabular}}} & 
\multicolumn{1}{l}{\multirow{2}{*}{\begin{tabular}[c]{@{}l@{}}Net-zero-energy\\Implimentation\end{tabular}}} 
\\
& & & Place & Activity & &\\ 
\hline \hline
2015 & \cite{lan2015estimating} & Piezoelectric & - & Walking/running & \color{red}\xmark & \color{red}\xmark \\
2017 & \cite{khalifa2017harke} & Piezoelectric & - & 5-activity & \color{red}\xmark & \color{red}\xmark \\
2017 & \cite{lan2017capsense, lan2020capacitor}& Piezoelectric & - & 5-activity & \color{red}\xmark & \color{red}\xmark \\
2018 & \cite{ma2018sehs, ma2020simultaneous} & Piezoelectric & - & Walking detection & \color{red}\xmark & \color{red}\xmark \\
2019 & \cite{aziz2019battery} & Radio frequency & 3D position & - & \color{red}\xmark & \color{red}\xmark \\
2019 & \cite{umetsu2019ehaas} & Solar cell, Piezoelectric & 9-place & - & \color{red}\xmark & \color{red}\xmark \\
2019 & \cite{sugata2019battery} & Solar cell & 8-place & - & \color{red}\xmark & \color{red}\xmark \\
2020 & \cite{sandhu2020towards} & Piezoelectric & 6-transport mode & - & \color{red}\xmark & \color{red}\xmark \\
2021 & \cite{Sandhu2021SolAREP} & Solar cell & - & 5-activity & \color{green}\cmark& \color{red}\xmark \\ 
2021 & ZEL & Solar cell, Piezoelectric & 8-place & Static/dynamic & \color{green}\cmark& \color{green}\cmark \\ 
\hline
\end{tabular}
\label{tab:related_work}
\end{table*}

\erase{
Considering the above background, it is clear that there is a need for wearable devices that use energy harvesters to convert ambient energy into electrical energy in order to power context recognition sensors and record the detected context information.
}

\add{
Considering the above background, our research group proposed Energy Harvesters As A Sensor(EHAAS)\cite{umetsu2019ehaas} and showed the vision of zero-energy lifelogging using a energy harvester. 
Energy harvesting is a technology that converts ambient energy into electrical energy and various application\cite{Winkel2020gameboy,fraternali2020ember,capharvester,maharjan2019high} using energy harvesting are proposed to extend the battery lifetime or make devices battery-less.
}

\erase{
For example, the power generated by a solar cell also provides information about the ambient light.
By using the energy harvester to collect data on the harvesting signal itself, the battery lifetime of the device can be extended to the point where the battery can be removed from the device. 
}

\erase{
In a previous study, Umetsu et al.\cite{umetsu2019ehaas} proposed the use of energy harvesters as a sensor (EHAAS) and showed that various energy harvesters, such as solar cells, piezoelectric elements, and Peltier elements, could be used as lifelogging sensors.
Later, Sandhu et al.\cite{Sandhu2021SolAREP} showed that zero-energy activity recognition is possible using solar cells as sensors. However, neither research has resolved the key challenges of implementing an energy harvesting function on a small wearable device that could then be used for achieving long-term lifelogging.
}
\add{
In a recent study, Sandhu et al.\cite{Sandhu2021SolAREP} showed that energy positive, i.e. zero-energy human activity recognition, including wireless communication is possible using solar cells as sensors.
However, neither research has resolved the key challenges of implementing an energy harvesting function on a small wearable device that could then be used for achieving long-term lifelogging.
}

In order to resolve this challenge, we propose a net-zero-energy lifelogging system named ZEL, which records when, where, and what activities office workers engage in on the office premises.
Net-zero-energy, which is a term used in relation to net-zero-energy buildings (ZEBs)\cite{ZeroEner14:online} and net-zero-energy houses (ZEHs), etc., means that the energy consumed by the overall system is covered by energy harvesting, thus achieving net-zero-energy consumption.
\erase{
More specifically, our goal was to design and implement a net-zero-energy lifelogging system that uses continuous energy harvesting to cover 99\% of the necessary power consumption.
}

\erase{
Our ZEL device, which uses two types of solar cells and a piezoelectric device as both power source and sensors, realizes net-zero-energy data collection by using a capacitor to record intermittent operations.
}
\add{
Our ZEL device, which uses two types of solar cells and a piezoelectric device as both power source and sensors to improve context recognition, realizes net-zero-energy data collection by using a capacitor to record intermittent operations.
}
In addition, seamless system state switching is achieved using a dual power switching mechanism, and \add{the} two comparators guarantee net-zero-energy and semi-permanent battery lifetime.
\erase{
The data, which are recorded in an onboard Electrically Erasable Programmable Read-only Memory (EEPROM) chip and extracted via Universal Serial Bus (USB) cable, are used by the application to build a lifelog that includes what activities a person has performed, where they were performed, and when.
}
\add{
The data recorded in the device is extracted via Universal Serial Bus (USB) cable and used to build a lifelog with a trained machine learning model.
}
In order to evaluate our ZEL system, we conducted data collection experiments in various environments.
The evaluation results show that an 8-place recognition accuracy level of 87.2\% (weighted F-measure) and \add{a} static/dynamic \erase{activities}\add{activity} binary classification accuracy of 93.1\% level (weighted F-measure) could be achieved.
In addition, practical testing showed that 99.6\% zero-energy operation, i.e., net-zero-energy operation, is possible.

The contribution\add{s} of this paper are summarized as follow:
\begin{itemize}
\item We designed and implemented the ZEL net-zero-energy lifelogging system \add{that uses heterogeneous energy harvesters to improve context recognition.}
\item In order to evaluate the proposed system, we conducted data collection experiments in different weather conditions (sunny, cloudy, and rainy) and on six different dates, during which we collected a total of 11 hours of data for 11 participants.
The evaluation results show that the PD model achieves an 8-place recognition accuracy level (weighted F-measure) of 87.2\% and a static/dynamic activities binary classification accuracy level (weighted F-measure) of 93.1\%.
\item By scheduling the system state using a dual power switching mechanism and two comparators, our proposed system was confirmed to work semi-permanently and achieve a 99.6\% zero-energy operation level in practical testing.
\end{itemize}

\erase{
The rest of this paper is organized as follows.
Section~\ref{sec:related_work} reviews work related to using energy harvesters as sensors, while Section~\ref{sec:proposed_system} presents our proposed system.
In Section~\ref{sec:implementation}, we describe the implementation of our proposed system.
Next, in Section~\ref{sec:eval_accuracy} and Section~\ref{sec:investigation}, we describe the evaluation results, the system's performance limits, and the zero-energy rate.
Finally, Section~\ref{sec:conclusion} concludes this paper.
}

\section{Related work} \label{sec:related_work}
This section describes the energy harvesting process and related work on context recognition using energy harvesters.

\erase{
In one such study, Ma et al.\cite{ma2019sensing} reported that energy harvesting materials are attracting attention in the field of large-scale deployment of wearable Internet of Things (IoT) devices due to issues such as battery life and disposal.
The authors also showed evidence that various wearable devices based on items such as backpacks\cite{xie2014human}, footwear\cite{zhao2014shoe}, and wristbands\cite{maharjan2019high}, as well as device applications\cite{Winkel2020gameboy,fraternali2020ember} using energy harvesting, are becoming increasingly popular.
In addition to the ability of energy harvesters to extend battery life, some studies have explored the use of energy harvesters as sensors for context recognition.
Some studies have explored the use of energy harvesters as sensors for context recognition.
}
\add{
Energy harvesting materials are attracting attention in the field of large-scale deployment of wearable Internet of Things (IoT) devices due to issues such as battery life and maintenance.
There are many studies of applications\cite{Winkel2020gameboy,fraternali2020ember, xie2014human, zhao2014shoe, maharjan2019high} using energy harvesters as a power source.
For example, WiWear\cite{WiWear} harvests RF energy from Wi-Fi transmissions and transmits accelerometer sensor data.
Apart from the approach of driving sensors by harvesting energy, there are efforts to use energy harvesters as context recognition sensors, taking advantage of the characteristics that harvesting energy is directly related to the ambient environment.
}
This field is particularly significant because using an energy harvester simultaneously as a sensor and a power source eliminates the need to install other sensors that would require power sources for context recognition, thus reducing the overall cost and power consumption of the device.
The following paragraphs describe some studies in which energy harvesters are used as sensors.

Khalifa et al.\cite{khalifa2017harke} reported on a performance evaluation of human activity recognition using kinetic energy harvesting (KEH), which is a process that converts kinetic energy into electric power, and showed that their proposed system consumes less power than conventional sensor-based systems.
Separately, Ma et al.\cite{ma2018sehs, ma2020simultaneous} proposed a mechanism to use KEH as a power source as well as a sensor, in which they consider the distortion of the harvesting energy's sensing signal during the energy harvesting process and proposed a filtering algorithm to compensate for it.
Then, using a device that implements their algorithm, they demonstrated that it could detect walking with higher accuracy than previous systems. 

Meanwhile, Lan et al.\cite{lan2015estimating} used KEH to classify walking/running and provided results that showed the classification accuracy of their system was close to that of an accelerometer.
However, that system did not have a harvesting function and thus required an external battery.
In another study, Lan et al. proposed CapSense\cite{lan2017capsense, lan2020capacitor}, which connects KEH to a capacitor and recognizes human activities based on its charging rate.
Then, using a wearable device with CapSense embedded in shoes, they demonstrated that their proposed system could achieve a 95\% accuracy level in recognizing five human activities while consuming 57\% less power than conventional systems.
However, during some static activities, the system took a long time to charge the capacitor. 

Aziz et al.\cite{aziz2019battery} proposed a zero-energy three-dimensional (3D) positioning system using the power of radiofrequency (RF) signals in which 64 antennas received radio waves generated by a \erase{huge} \add{lots of} beacon\add{s}, thereby enabling highly accurate 3D positioning.
However, their proposal could not be considered a net-zero-energy system because it requires significant amounts of power on the environmental side.
Meanwhile, Umetsu et al.\cite{umetsu2019ehaas} proposed an EHAAS-based\erase{,} room-level place recognition system for lifelogging in which they showed that that solar cells are the best currently available harvesters for room-level place recognition and could achieve highly accurate place recognition with only two types of solar cells, while Sugata et al.\cite{sugata2019battery} implemented EHAAS on a nametag-shaped device and achieved highly accurate place recognition under limited conditions. 
Sandhu et al.\cite{sandhu2020towards} used KEH to recognize transportation systems such as trains, ferries, etc., from the received vibrations.
After large-scale data collection experiments, they reported that some transportation systems with large vibrations, in which the power generated by the energy harvester exceeded the power used to record sensor signals, had achieved zero-energy consumption.
In a separate study, Sandhu et al.\cite{Sandhu2021SolAREP} used solar energy harvesting (SEH), which converts solar energy into electricity, to recognize five human activities.
That system incorporates a human activity recognition pipeline, including everything from sensor signal acquisition to wireless communication, in a wearable device, and the authors showed that the pipeline could be executed with net-zero-energy consumption. However, the prototype used for the evaluation still required an external battery because it did not have a harvesting function.

Each of the related works discussed above is summarized in TABLE~\ref{tab:related_work}.
Here, it should be noted that although many of those studies aimed at reducing the power consumption of context recognition, they required the use of external power sources and thus were not net-zero-energy systems.
For example, some studies reported systems that achieved zero-energy in limited environments, such as while walking, but were not net-zero-energy because they relied on battery power at other times.
Additionally, some studies have shown that net-zero-energy can be achieved by measuring the power consumption of the harvesting energy and the overall system, but none have been implemented to date.
\erase{In order to} \add{To} achieve high accuracy and reliable operations in various environments, we employed two harvesters with different---stable and unstable---power generation capabilities.
The stable harvester guarantees a wide operational environment, while the unstable harvester enables highly accurate context recognition.
The latter is significantly affected by the surrounding ambient.
\erase{
In addition, two comparators are used to minimize battery use, thus resulting in a net-zero-energy device.
}
\add{Note that wireless transmission is not included in our study.}

\section{ZEL: net-zero-energy lifelogging system using heterogeneous energy harvesters} \label{sec:proposed_system}
\begin{figure}[bt]
 \centering
 \includegraphics[width=0.9\hsize]{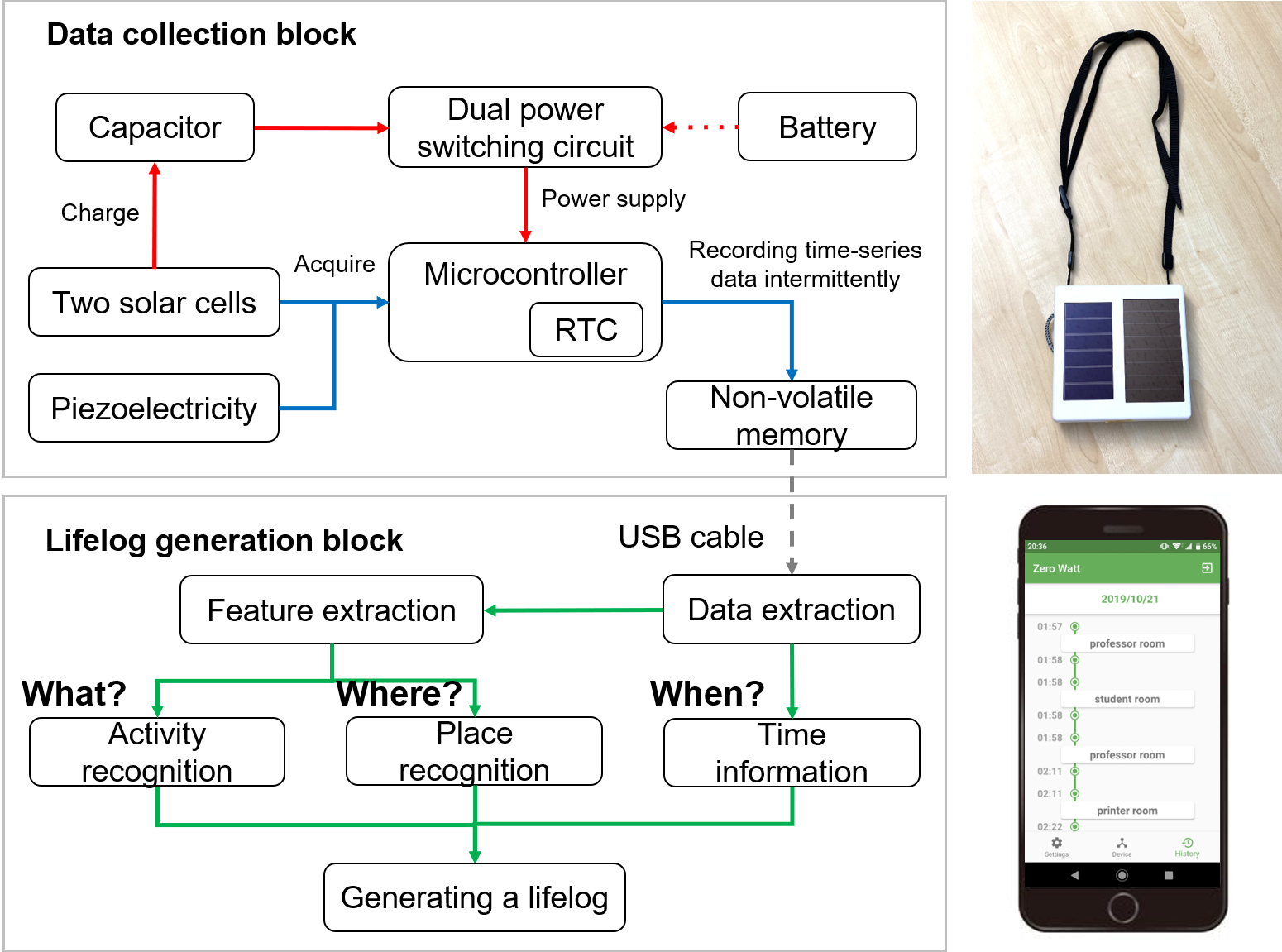}
 \caption{Proposed ZEL model consisting of data collection and lifelog generation blocks.}
 \label{fig:system_configuration}
\end{figure}

Our ZEL system consists of energy harvesting\add{,} data collection\add{,} and lifelog generation blocks, as shown in Fig.~\ref{fig:system_configuration}.
\add{
It might be questionable that the system has a battery despite net-zero-energy, but it is only used for time keeping.
Hence, no need for battery maintenance or recharging.
}
In the data collection block, the self-powered, nametag-based wearable device records lifelogging data intermittently.
In the lifelog generation block, the data are manually extracted from the device, and the location and activities of the wearer are recognized.
Next, the block builds a lifelog containing three types of information: when, where, and what activities the wearer engaged in.
In this section, we describe the operations of each block.

\begin{figure}[bt]
 \centering
 \includegraphics[width=0.9\hsize]{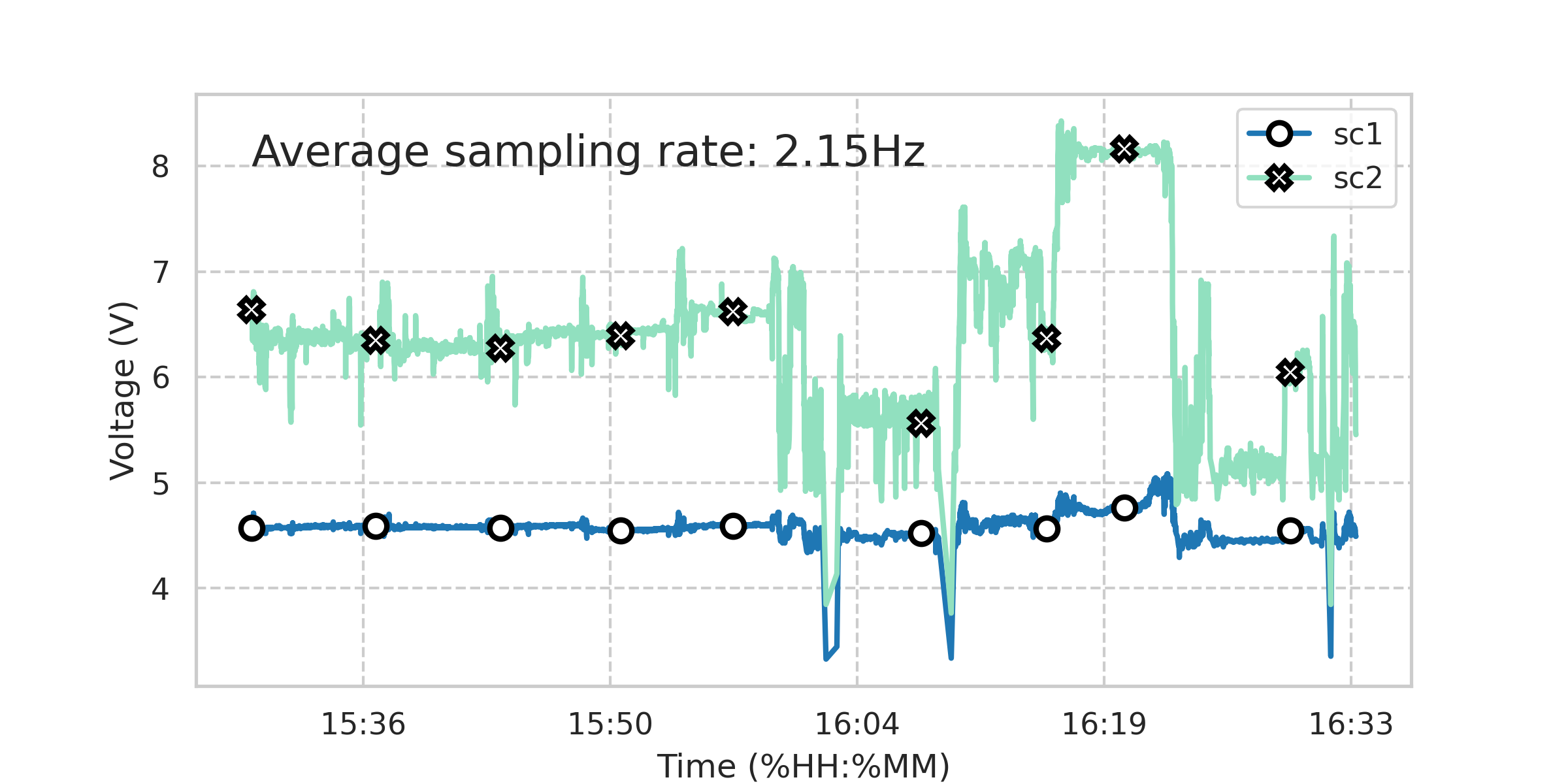}
 \caption{Transition of solar cell's \erase{harvesting energy(open-circuit)} \add{voltage} when visiting various locations (average sampling rate: 2.15 Hz).} 
 \label{fig:voltage_sensor_data}
\end{figure}
\begin{figure*}[bt]
 \centering
 \includegraphics[width=0.80\hsize]{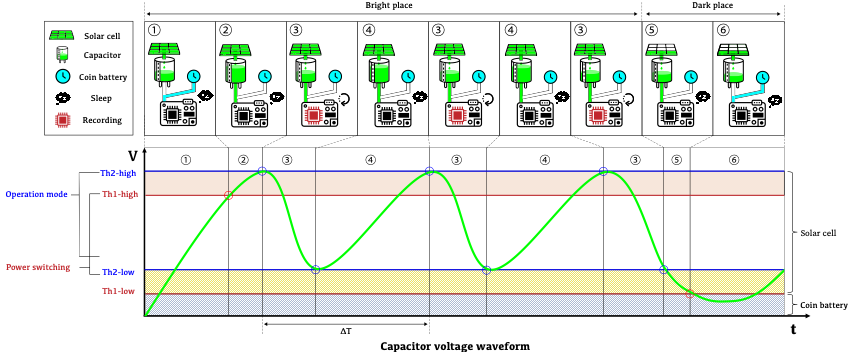}
 \caption{System state scheduling by capacitor voltage using two \erase{voltage detectors} \add{comparators} (when the user moves from bright to dark locations).}
 \label{fig:device_flow}
\end{figure*}
\subsection{Data collection block}
In the following subsections, we first explain the concept behind achieving self-powered data collection operation that works by recording the harvesting signal, beginning with a description of the system state\erase{s} scheduling using two \erase{voltage detectors} \add{comparators}.

\subsubsection{Recording time-series data \add{by} using harvesting energy}
Inspired by the concepts presented by Sugata et al.\cite{sugata2019battery} and Capsense\cite{lan2017capsense, lan2020capacitor}, we explore the process of using energy harvesting data to record time-series information.
By connecting the energy harvester\erase{s} to a capacitor, the microcontroller is activated intermittently.
In other words, when the microcontroller is activated, it acquires the \erase{harvester signal(open-circuit voltage),}\add{harvester's voltage} and the capacitor is recharged when the microcontroller sleeps.
\erase{In order to} \add{To} recognize location\add{s} and activities, we employ the same heterogeneous solar cells and piezoelectric elements that previous studies have confirmed function effectively for place and activity determinations\cite{umetsu2019ehaas}.
We combined one dye-sensitized solar cell, which has stable power generation characteristics even indoors, and one amorphous solar cell, which has unstable power generation characteristics, to enable the stable operation and high accuracy context recognition.

The location and activity data are obtained by these harvesters, while time time-specific data are obtained by the real-time clock (RTC).
Since RTCs need to retain time even when the system is down, the device needs a timekeeping battery.
A de-facto standard solution adopted by many mobile devices is integrating a small timekeeping battery into the circuit board, and we follow this approach for our system.
By recording time data together with harvester signals, time-series data are created and then written to non-volatile memory.
By integrating all of the above steps, we could design and implement a wearable device that can semi-permanently operate at net-zero-energy consumption levels using its own energy harvesting system, thus eliminating the need for battery \erase{recharging} \add{maintenance}.

Fig.~\ref{fig:voltage_sensor_data} shows the power generated when wearing the ZEL system device and visiting various locations.
As can be seen in the figure, the voltage from the dye-sensitized solar cell (sc1) is generated stably at all locations while \add{the} power generation by the amorphous solar cell (sc2) fluctuates significantly.
\add{
These solar cells enable both a stable power supply and high recognition accuracy.
}
The average sampling rate in our real-world operation tests was about 2.15 Hz.


\begin{figure}[t]
 \centering
 \includegraphics[width=\hsize]{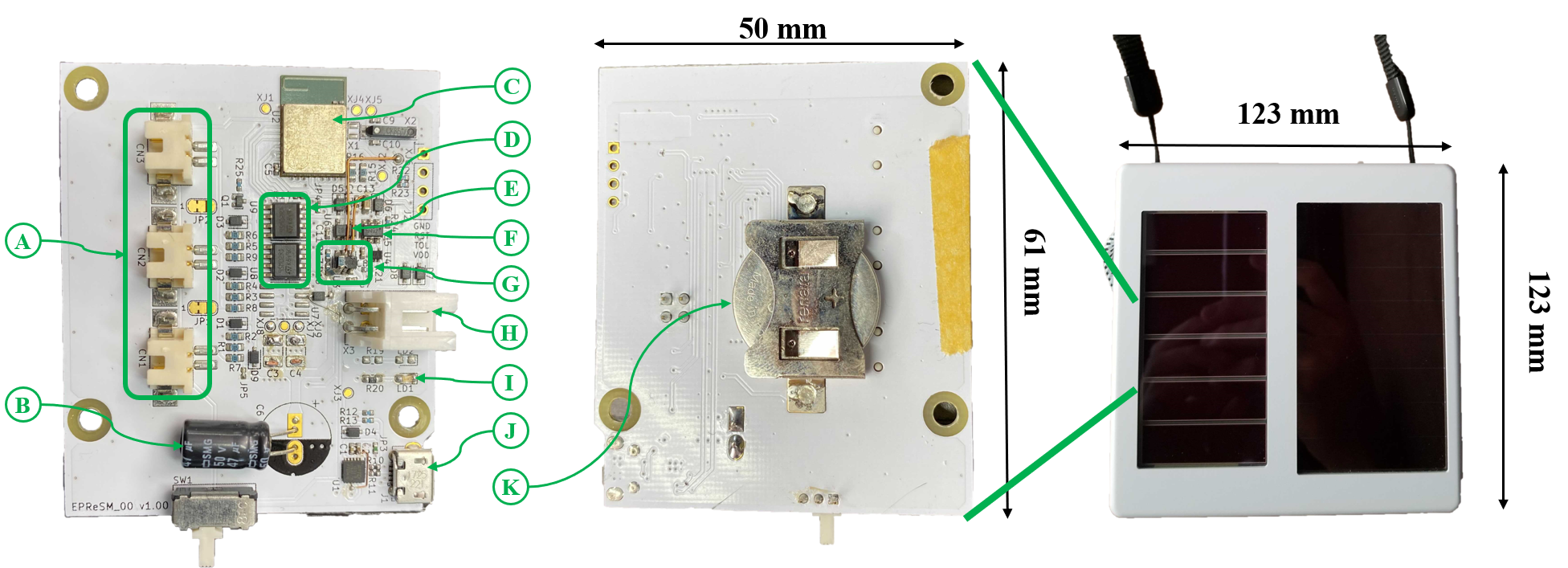}
\caption{Primary ZEL system components: (A) solar cells connectors, (B) a 47 $\mu$ F\,capacitor, (C) a Renesas RY7011 BLE module with built-in RL78/G1D and ultra-low power antenna, (D) a STMicroelectronics M95256-WMN6P SPI bus EEPROM, (E) a Ricoh RP118N221B-TR-FE low-dropout (LDO) voltage regulator, (F) a Texas Instruments TPS22860DBVR ultra-low leakage current load switch, (G) XC6134C21EMR-G and XC6134C22CMR-G \erase{voltage detectors} \add{comparators}, (H) a piezoelectricity connector, (I) an LED, (J) a Micro-USB connector, and (K) a  CR2032 coin battery.} 
 \label{fig:circuit}
\end{figure}

\subsubsection{System state scheduling using dual power switching mechanism and two \erase{voltage detector} comparators}
\erase{In order to} \add{To} guarantee semi-permanent, net-zero-energy consumption, the device must:
\begin{itemize}
 \item be able to recover after a visit to a dark location
 \item be able to minimize battery use
\end{itemize}
We satisfied these requirements by adopting system state scheduling using a dual power switching mechanism and two \erase{voltage detectors} \add{comparators}.

Intermittent operation via energy harvesting is usually implemented by \erase{voltage detectors} \add{comparators} with hysteresis characteristics\cite{electronics8121446,Chen2017BatteryfreePM}.
The \erase{voltage detector} \add{comparator} input is connected to a capacitor, and the output switching occurs when the capacitor is charged by energy harvesters or discharged by system power consumption.
The system becomes active when the \erase{voltage detector} \add{comparator} detects that the capacitor has been charged with sufficient power to operate the system and enters the sleep mode to recharge the capacitor when the detector senses power consumption.
However, in situations where the ZEL system is unable to generate sufficient power (such as when visiting a dark place), the power switching circuit needs the ability to switch to battery power consumption \erase{in order} to keep operating.
Since battery consumption needs to be minimized to achieve net-zero-energy, we added a state scheduling mechanism that uses an additional \erase{voltage detector} \add{comparator}.

One of the two \erase{voltage detectors} \add{comparators} is for switching the operation mode (sleep or active) of the microcontroller, while the other is for power switching (capacitor or battery).
Both \erase{voltage detectors} \add{comparators} have hysteresis characteristics \erase{\erase{voltage detectors} \add{comparators}} and each has two threshold values.
The detector thresholds for the operating mode are a positive offset of the power switching thresholds.
Fig.~\ref{fig:device_flow} shows the system state transition with capacitor voltage.
In Fig.~\ref{fig:device_flow}, states 1 to 4 show \add{the} operation in a bright location, and states 5 and 6 show \add{the} operation in a dark place.
Details of each state are shown below:

\begin{enumerate}
 \renewcommand{\labelenumi}{\textcircled{\scriptsize \theenumi}}
 \item Power is supplied from the battery, and the microcontroller is in sleep mode.
 \item The power source switches to the capacitor.
 \item The microcontroller transits to the active mode and starts the recording. After recording, power is quickly consumed by the light-emitting diode (LED).
 \item The microcontroller transits to the sleep mode, and the capacitor is charged if sufficient power generation is available.
 \item The microcontroller transits to the sleep mode, and the capacitor voltage decreases if power generation is insufficient.
 \item The power source switches to the battery.
\end{enumerate}

As shown in the above state scheduling, the use of two \erase{voltage detectors} \add{comparators} enables power supply from the capacitor even when the capacitor is being charged, thereby minimizing battery use.

\subsection{Lifelog generation block}
Using the time-series data recorded in the data collection block, the lifelog generation block generates lifelog data from three information types: when, where, and what activity.
Where and what activities are predicted using machine learning (ML).
In the following section, we describe each lifelog generation block operation (prepossessing, feature extraction, and ML) in greater detail.

\subsubsection{Prepossessing}
First, since the data collected immediately after power-on may contain outliers, the first 30 seconds are removed.
Next, the sampling rate is added to the data calculated from the timestamp acquired by the RTC because it has also been confirmed that the rate of intermittent operation is important for context recognition\cite{lan2017capsense,lan2020capacitor}.
In order to apply time-related features, each data type is divided \add{by} a fixed-length window.
In our system, we selected a window size of 1.24 seconds\cite{torigoe2020strike, nakamura2019waistonbelt}, which is commonly used in accelerometer activity recognition.
The window overlap rate is 50\%.

\subsubsection{Feature extraction}
Since time-series data are the target of this study, we use the following 17 features that have been validated in previous studies\cite{torigoe2020strike,nakamura2019waistonbelt} related to activity recognition using accelerometers: mean, standard deviation, median absolute deviation, maximum, minimum, sum of squares, entropy, interquartile range, fourth-order Burg autoregressive model coefficients, range of minimum and maximum values, root mean square, frequency signal skewness, frequency signal kurtosis, maximum frequency component, frequency signal weighted average, frequency band spectral energy, and power spectral density.

\subsubsection{Machine learning}
To construct an ML model that outputs place labels and activity labels, we examined nine popular machine learning algorithms (support vector machine (SVM), artificial neural network (ANN), random forest (RF), decision tree (DT), Light Gradient Boosting Machine (LightGBM), logistic regression (LR), K-nearest neighbor (KNN), Naive Bayes (NB), and extra-trees (ET)) to determine the one most suitable for classification and embedding in the final proposed system.
Since lifelogging does not require detailed position information, we perform majority voting for a certain number of samples in relation to each place and activity predicted by the model.
In the next section, we will show how the accuracy of majority voting depends on the number of samples (1 sample: 1.24 seconds).

\section{ZEL implementation} 
\label{sec:implementation}
We implemented the design described in Section~\ref{sec:proposed_system} on a printed circuit board and embedded it in a nametag-shaped wearable device based on the expectation that it would be used by office workers.

\subsection{Circuit\add{s} and wearable device}
The ZEL circuit was implemented as shown in Fig.~\ref{fig:circuit}, where it can be seen that the circuit is 50 $\times$ 61 mm in size, weighs 15 grams, and contains three solar cell connectors and one piezoelectric connector.
The nametag-shaped device into which the circuit is embedded is 123 mm square on each side and weighs 192 grams, so it can be comfortably worn around the user's neck without interfering with desk work \erase{, as shown in Fig.~\ref{fig:ZEL}}.
Renesas RY7011 is used as the microcontroller to achieve ultra-low power consumption performance.
We use a capacitor connected to the harvester\erase{s} as the main power source and a coin battery as the supplementary power source.
The proposed scheduling mechanism described in Section~\ref{sec:proposed_system} minimizes the use of a coin battery to achieve net-zero-energy.
The collected data are first stored in EEPROM IC, and then passed to the lifelog generation block via \erase{Micro} \add{micro} USB cable.

\subsection{ZEL energy harvesters}
As described in Section~\ref{sec:proposed_system}, we employ one dye-sensitized solar cell (sc1), one amorphous solar cell (sc2), and one piezoelectric element as energy harvesters.
The specifications of each are shown in TABLE~\ref{tab:specification_of_harvesters}.
Different solar cells type were used to ensure a wide operating range and high recognition accuracy, as described in Section~\ref{sec:related_work}.
A weight consisting of a screw and nut is attached to the tip of the piezoelectric element, which detects \add{the} vibrations generated by user activity.
Since we did not consider the use of the piezoelectric element as a power source when designing the circuit, this device only uses \add{the} power obtained from the solar cells.
The two solar cells are connected in parallel to a capacitor, and the harvesting signal is acquired by connecting those voltages to the microcontroller Analog/Digital (A/D) port.

\begin{table}[bt]
 \centering
 \caption{ZEL energy harvesters specifications: sc1, sc2, and piezoelectric element.}
 \begin{tabular}{cccc} \hline
 Type & Dye-sensitized (sc1) & Amorphous (sc2) & Piezoelectric \\ \hline \hline
 Image & 
 \begin{minipage}{20mm}
 \centering
 \includegraphics[width=0.6\hsize]{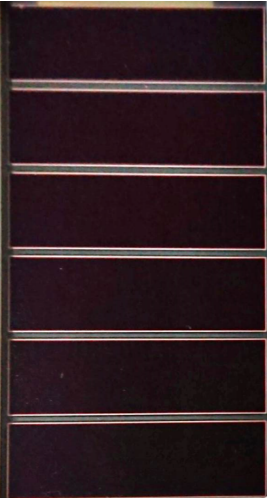}
 \end{minipage} &
 \begin{minipage}{20mm}
 \centering
 \includegraphics[width=0.6\hsize]{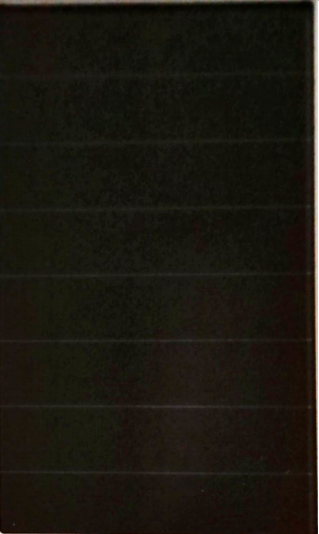}
 \end{minipage} &
 \begin{minipage}{7mm}
 \centering
 \includegraphics[width=0.6\hsize]{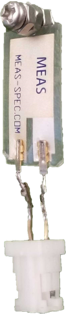}
 \end{minipage} \\ \hline
 Power & 252 $\mu$W & 332 $\mu$W & 400 mV/g \\ \hline
 Size & 97 $\times$ 57 mm & 96 $\times$ 47 mm & 13 $\times$ 25 mm \\ \hline
 Weight & 20.0 g & 16.2 g & 2.5 g \\ \hline
 \end{tabular}
 \label{tab:specification_of_harvesters}
\end{table}

\subsection{Implementation of system state scheduling}
\begin{figure}[t]
 \centering
 \includegraphics[width=1.0\hsize]{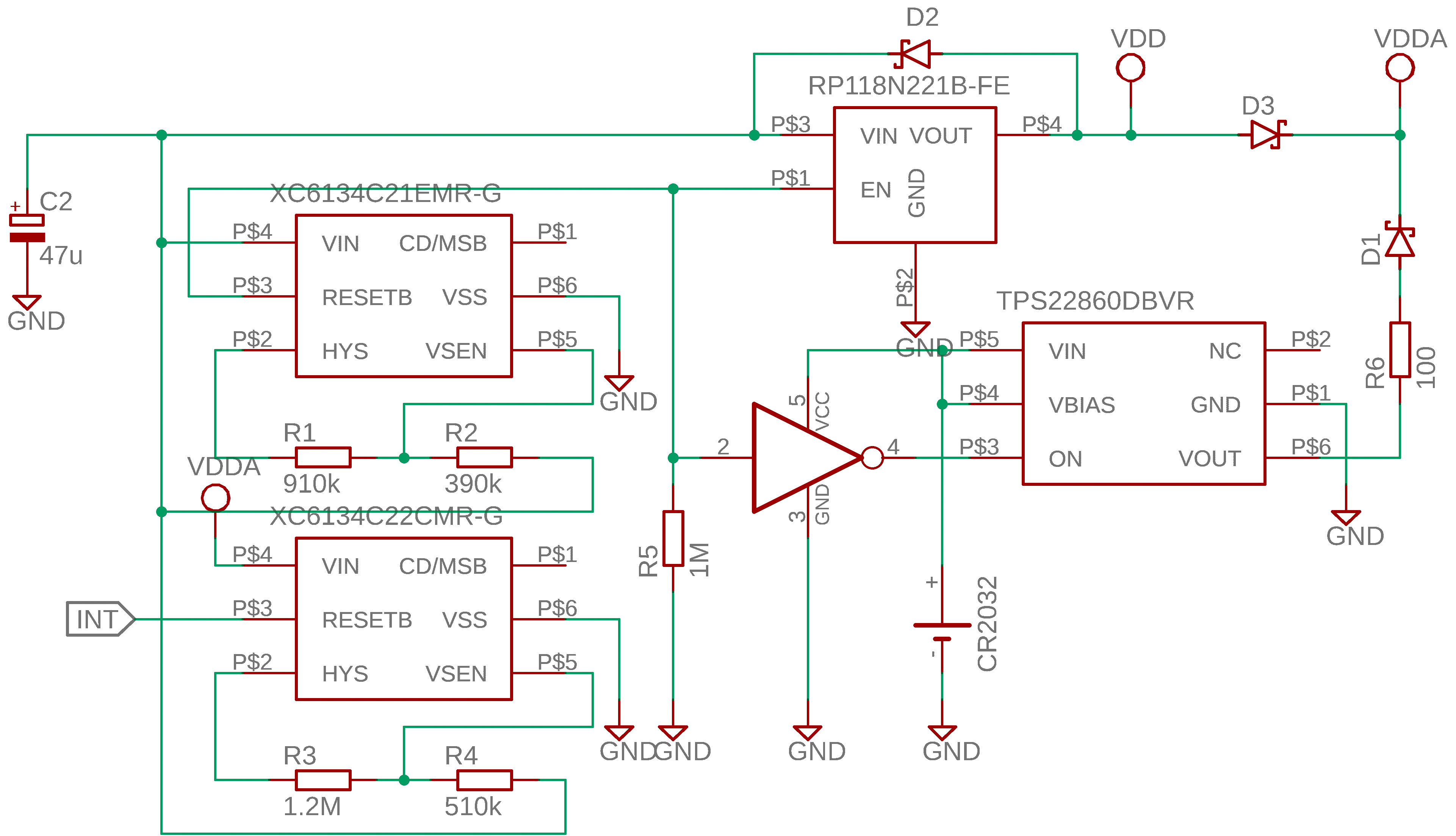}
 \caption{Schematic of the system state scheduling component (some ceramic capacitors are omitted for simplicity).}
 \label{fig:schematic}
\end{figure}

System state scheduling is implemented using two \erase{voltage detectors} \add{comparators}, an LDO voltage regulator, and a load switch, as shown in Fig.~\ref{fig:schematic}.
The \erase{voltage detector} \add{comparator} output for power switching is connected to the LDO ENABLE (EN) pin, and the inverted output is connected to the ON pin of the load switch to which the coin cell battery is connected, thus realizing the dual power switching mechanism.
The \erase{voltage detector} \add{comparator} output used for operation mode switching is connected to the microcontroller interrupt pin, which operates as a trigger to switch between the microcontroller sleep and active states.
This setup enables system state scheduling, as shown in Fig.~\ref{fig:device_flow}.

\begin{figure}[t]
 \centering
 \includegraphics[width=\hsize]{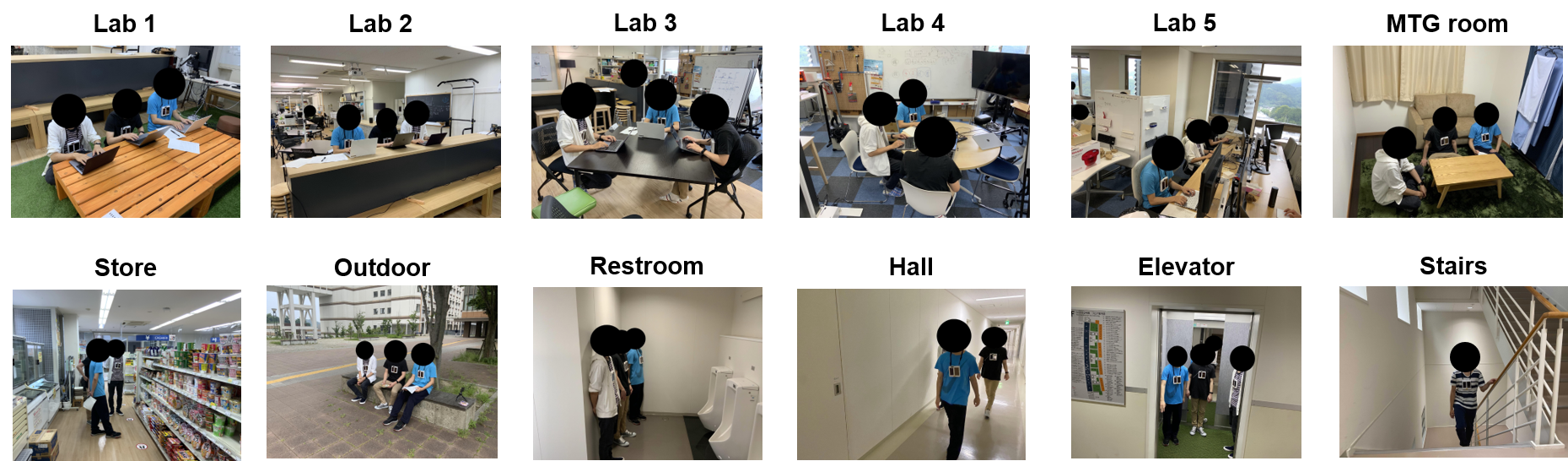}
 \caption{Scenes from data collection experiment.}
 \label{fig:experiment}
\end{figure}

\section{Evaluating ZEL recognition accuracy } \label{sec:eval_accuracy}
In this section, we describe the data collection method and the experimental environment used for evaluating our proposed system, after which we evaluate the following three items:

\begin{itemize}
 \item Scaling the number of majority samples
 \item Comparing previous studies and sensor-based methods
 \item Evaluating person-independent (PI) model versatility
\end{itemize}
Weighted F-measures were used as the evaluation metrics.

\subsection{Data collection and experimental environment}
This subsection reports on data collection experiments conducted to evaluate our ZEL system.
Here, it is important to note that since solar cells are used as energy harvesters, the influence of external light should always be considered.
Therefore, we conducted data collection experiments with the assistance of 11 participants during different weather conditions (sunny, cloudy, and rainy) on six different dates.
In order to ensure practicality, we made a list of possible activities and places to visit in a day at our university and designed a scenario that covers 14 locations and five activities.
\erase{
, as shown in TABLE~\ref{tab:scenario}.
The participants stay and move as indicated from the top to the bottom of this table. 
}
\add{
The participants stay and move from one place to another.
}

To facilitate comparison\erase{s} with the conventional method, the participants wore a ZEL device, an accelerometer, and an illumination sensor and were asked to go about their activities according to the experimental scenario.
In order to facilitate data acquisition, there were some locations where we specified participant activities.
For example, in the laboratory, the participants sit and work.
In summary, we collected data on 14 locations and five activities: sitting, standing, walking, and traveling upstairs/downstairs for the 11 participants\add{,} covering a total of 11 hours under various lighting conditions.
An overview of the data collection experiment is shown in Fig.~\ref{fig:experiment}.
We then manually annotated the collected data to indicate the places and activities.
Since detailed position information is not particularly important for lifelog compilation, \add{the} rough locations, labs 1 to 5, and the hallways of each floor were grouped together.
As for activities, since it is important to know how much exercise they performed during the day, activities were grouped into two categories: static (sitting, standing) and dynamic (walking, traveling upstairs/downstairs).
\add{
The collected data is downloaded via USB and the ML model is trained and applied offline.
}

\subsection{Scaling labels via majority voting}
As described in Section~\ref{sec:proposed_system}, we apply majority voting to the labels output by the model.
Majority voting also plays a role in removing noise, particularly in terms of location, because transitions are often slow.
Accordingly, \erase{in order} to determine the optimal number of majority samples, we recognize multiple numbers of majority samples and perform evaluations with the PD model obtained by 10-fold cross-validation (CV) for each user.
The ML algorithms with the highest average accuracy, which are LightGBM for place recognition and SVM for activity recognition, were used in our evaluation\erase{s}.
The average user accuracy levels for place and activity recognition are shown in Fig.~\ref{fig:majority_sample_scaling} for each sample size.
Our obtained results showed that the best accuracy was acquired for a majority sample size of 20 (13.02 seconds), which proved an approximately 2\% accuracy improvement over cases without majority voting.
Therefore, we used a sample size of 20 in the following evaluation.


\subsection{Comparison \erase{to} \add{with} previous studies and sensor-based methods}
\begin{figure}[bt]
 \centering
 \includegraphics[width=0.90\hsize]{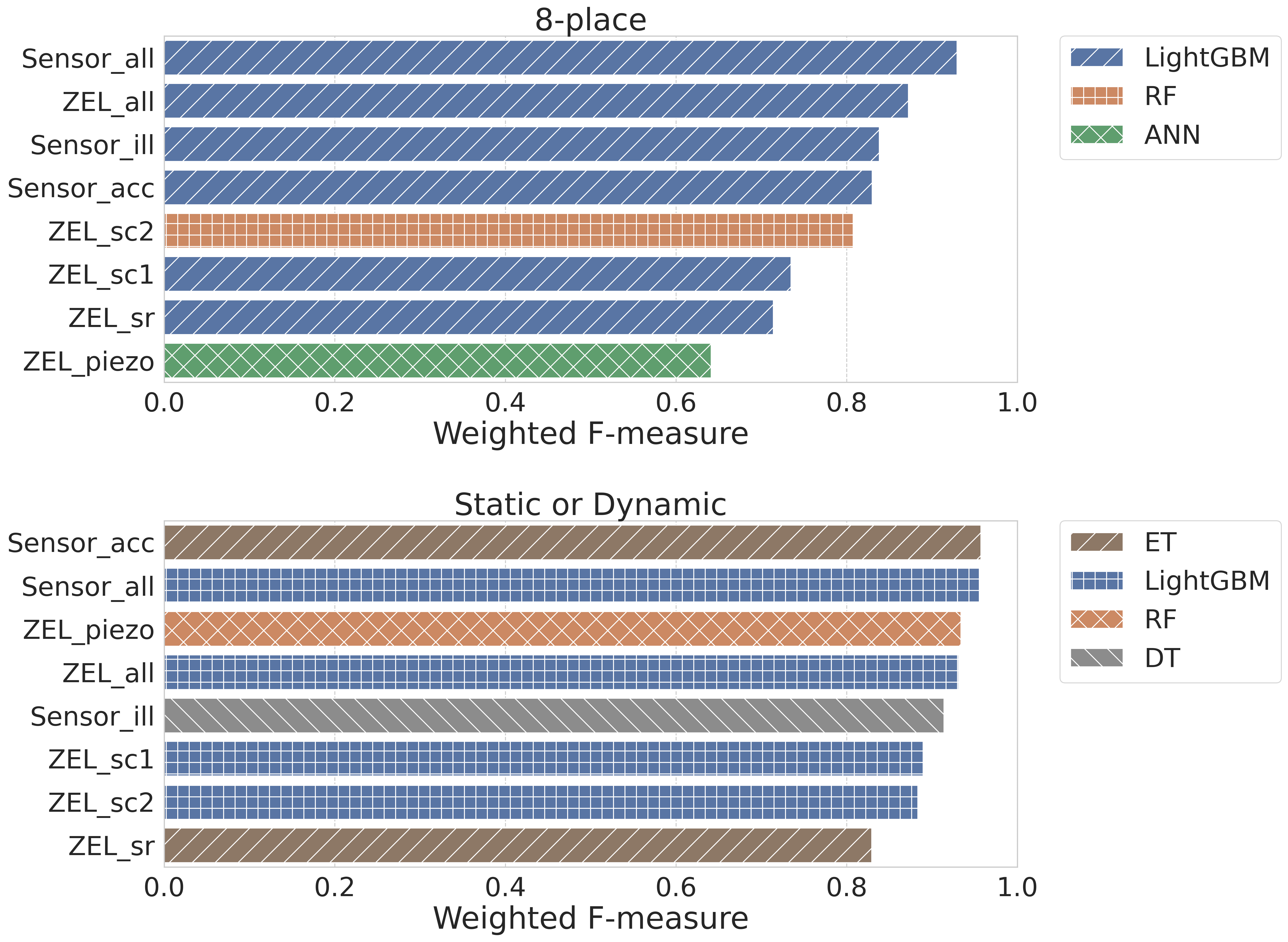}
 \caption{Accuracy of 8-place and static/dynamic activity recognition for each method.}
  \vspace{-5mm}
 \label{fig:10fold_accuracy}
\end{figure}

In order to compare previous studies and conventional sensor-based methods using an accelerometer (acc) and an illuminance sensor (ill), we performed 8-place recognition summarizing the laboratory and hallway labels and classified static/dynamic activities.
As in the previous subsection, we used the PD model for the evaluation.
To ensure a fair comparison, the acc up\erase{-}sampling rate was set at 100 Hz and applied to the data acquired by both the ZEL and the illuminance sensor.
We used linear interpolation for up-sampling.
The results obtained by choosing the most accurate model are shown in Fig.~\ref{fig:10fold_accuracy}. 

The models with underscores represent the data used.
For example, \textit{sc1} refers to the amount of energy harvested by the dye-sensitized solar cell, \textit{sc2} refers to the amount of energy harvested by the amorphous solar cell, \textit{sr} means the sampling rate, \textit{acc} means the  accelerometer, and \textit{ill} means the illumination sensor.
Taken further, \textit{ZEL\_sc1} refers to a model created only from the dye-sensitized solar cell power harvesting data and does not include data from the piezoelectric element.
In contrast, \textit{Sensor\_all} is a model created using both accelerometer and illumination sensor data.
Since the ML algorithm hyperparameters were not adjusted, their default values were used in all cases. 

Looking at the results obtained, we found that for place recognition, ZEL achieved the second-highest accuracy after the model combining the accelerometer and illuminance sensor.
This accuracy level is comparable to that of previous studies\cite{umetsu2019ehaas,sugata2019battery}, thereby indicating that the ZEL device is capable of place recognition under practical usage conditions.
In most cases, the LightGBM accuracy was the highest.
This may be due to PD model overfitting.
For activity recognition, the model using only piezoelectric element data achieved the highest accuracy rating of 93.2\%, followed by the model using the accelerometer.
\erase{
It was also confirmed that the models using only the solar cell harvesting signals (ZEL\_sc1 and ZEL\_sc2) could achieve sufficient accuracy when used in binary classification.
}
\add{
It was also confirmed that the model that combined multiple harvesters achieve better accuracy in place recognition.
}
This is due to the fact that the harvesting signals changed as a result of shadows created during human activities.
These comparisons show that the ZEL system can recognize places and activities with an accuracy level close to conventional sensors that consume significant amounts of power.

\subsection{Versatility evaluation}
\begin{figure}[bt]
 \begin{tabular}{cc}
 \begin{minipage}[t]{0.45\hsize}
 \centering
 \includegraphics[width=1.0\hsize]{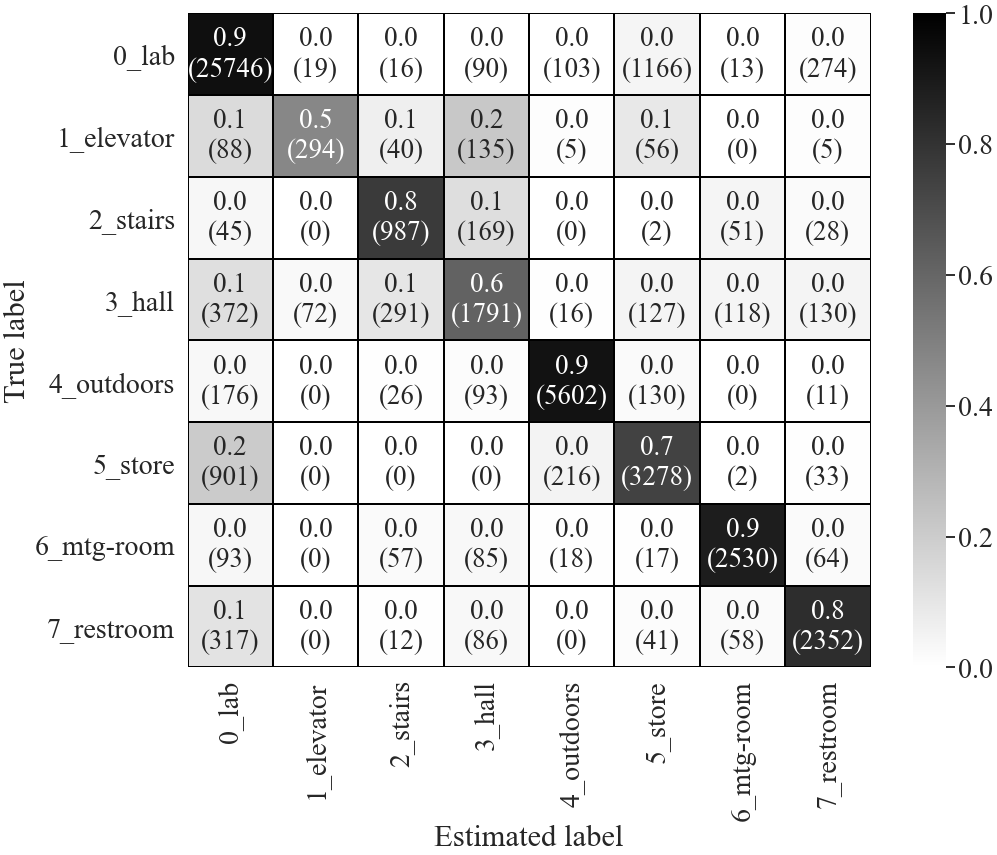}
 \subcaption{8-place (PD)}
 \label{fill}
 \end{minipage} &
 \begin{minipage}[t]{0.45\hsize}
 \centering
 \includegraphics[width=1.0\hsize]{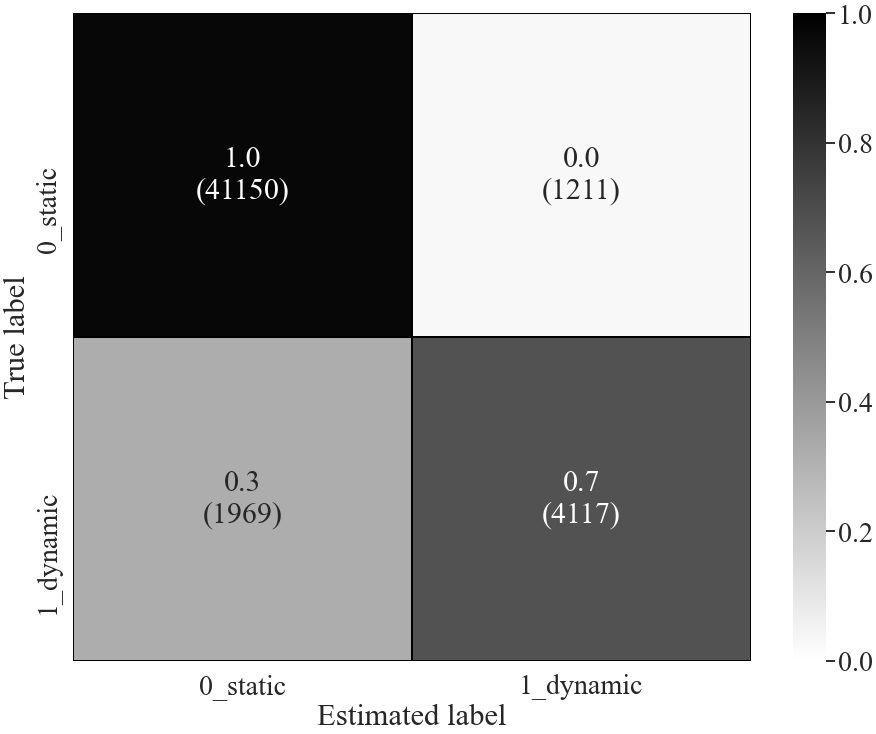}
 \subcaption{static/dynamic (PD)}
 \label{transform}
 \end{minipage} \\
 
 \begin{minipage}[t]{0.45\hsize}
 \centering
 \includegraphics[width=1.0\hsize]{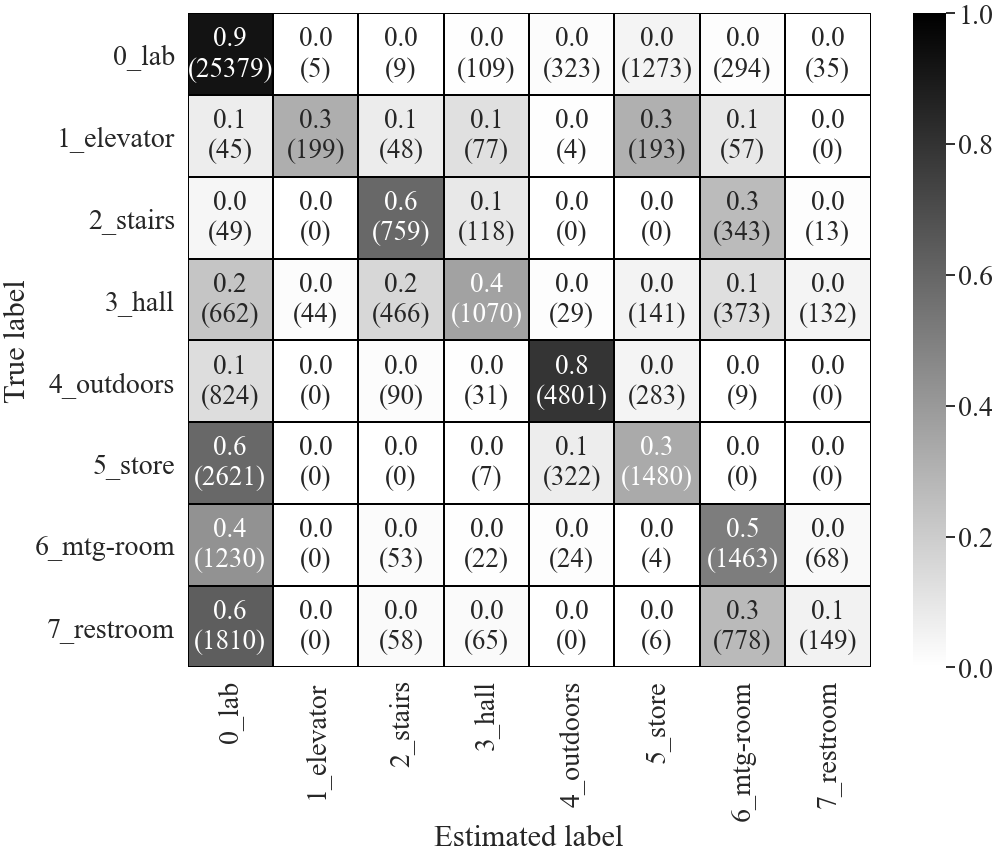}
 \subcaption{8-place (PI)}
 \label{LOUO_place}
 \end{minipage} &
 \begin{minipage}[t]{0.45\hsize}
 \centering
 \includegraphics[width=1.0\hsize]{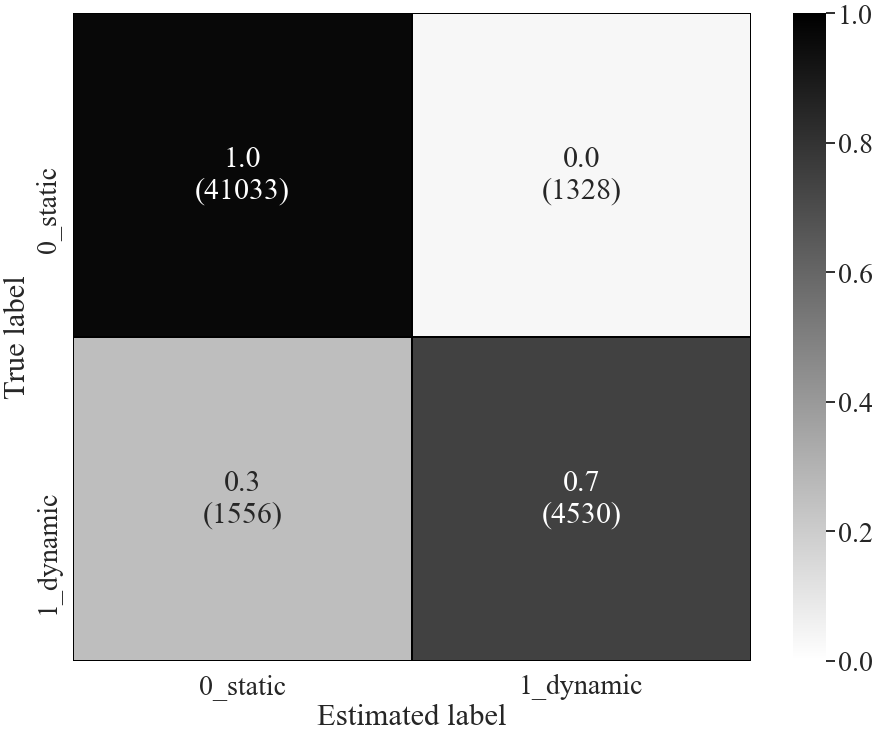}
 \subcaption{static/dynamic (PI)}
 \label{LOUO_action}
 \end{minipage}
 \end{tabular}
 \caption{Confusion matrix for each recognition obtained by PD (10-fold CV) and PI (LOUO CV) models.}
 \vspace{-2mm}
 \label{fig:cv}
\end{figure}

Next, we compared the PD model results with those from the PI model obtained via leave-one-user-out (LOUO) CV \erase{in order} to evaluate the versatility of ZEL \erase{in reference to} \add{regarding} the participants.
The confusion matrix acquired by the PD and PI models is shown in Fig.~\ref{fig:cv}.
While both models show high accuracy in classifying activities, the PI model shows significantly lower place recognition accuracy.
In particular, the recognition accuracy for restrooms was about 10\%. This may be due to the fact that the number of light sources in the restroom is small and because the amount of energy harvested varies significantly depending on the user's position.
Other factors, such as the participant's height and posture, are also considered.

\section{Investigation of performance limits and zero-energy rate} \label{sec:investigation}
In this section, we describe the performance limits and the zero-energy rate of our ZEL system in practical use.

\subsection{Performance limits}
To investigate the ZEL system performance limits, we performed 14-place and 5-activity recognition with subdivided labels.
The result of each CV is summarized in TABLE~\ref{tab:cv_for_each_target}.
The accuracy of the 14-place recognition in the PI model was reduced by about 32\% compared to 8-place recognition.
Fig.~\ref{fig:14places_cm} shows the confusion matrix of the 14-place recognition acquired by the PI model.
The significantly lower accuracy levels between labs and hallways indicate that ZEL is unsuitable for recognizing fine place details within the same room or in a similar lighting environment.
As for 5-activity recognition, there were many misrecognitions for walking and traveling upstairs/downstairs.
This indicates that it is necessary to reconsider the position of the piezoelectric element to obtain more detailed activity recognition because numerous previous studies, such as \cite{ma2018sehs}, report achieving high action recognition level levels using piezoelectricity.

\begin{table}[bt]
 \centering
 \caption{Results of recognition accuracy for each classification target and each PD and PI model.}
 \label{tab:cv_for_each_target}
 \begin{tabular}{lll} \hline
 Target & PD & PI \\ \hline \hline
 8-place & 0.873 (LightGBM) & 0.728 (LR) \\
 static/dynamic & 0.932 (LightGBM) & 0.943 (LightGBM) \\
 14-place & 0.832 (LightGBM) & 0.497 (LR) \\
 5-activity & 0.821 (LightGBM) & 0.728 (SVM) \\
 \hline
 \end{tabular}
\end{table}


\subsection{Zero-energy rate}
As shown in Fig.~\ref{fig:device_flow}, the ZEL device cannot operate and record lifelog information in dark places because it cannot generate sufficient power.
Therefore, we investigated how much of the system activity (i.e., zero-energy rate) occurs when the ZEL device is used in a real environment.
For this investigation, we connect the load switch output voltage (VOUT) of the TPS22860DBVR in Fig.~\ref{fig:schematic} to the data logger (Adafruit Feather M0 Adalogger).
This pin output becomes HIGH when the system is inoperable, i.e., not zero-energy.
Hence, by recording this output to the SD card in the data logger, the zero-energy rate can be determined.

Next, we conducted an experiment in which one participant wore the ZEL device with a data logger to work from 10:00 to 19:00 without any place or activity restrictions. The experiment was conducted for a total of four days, and a total of 36 hours of data were acquired.

To show that the proposed method minimizes the battery usage, we compare the ZEL device output with the zero-energy rate of a circuit version containing only one \erase{voltage detector} \add{comparator} that did not minimize battery usage.
In the results shown in TABLE~\ref{tab:activation_ratio}, it can be seen that the proposed method improved the zero-energy rate and the average not zero-energy rate was less than 1\%.
\erase{
It is worth mentioning that the zero-energy rate is just a time related rate not power consumption rate.
}
However, about 1\% of the time when the battery is in use is state 6 in Fig.~\ref{fig:device_flow}, and the microcontroller is in sleep mode, so \add{the} power consumption is very low.
From this, it can be said that the ZEL device achieved net-zero-energy.

\begin{table}[bt]
\centering
\caption{Zero-energy rate of the system in practical use (average of four days). The baseline method is using only one \erase{voltage detector} \add{comparator} and the proposed method is using two \erase{voltage detectors} \add{comparators} to minimize battery usages.}
\begin{tabular}{lrrrr}
\hline
\multirow{2}{*}{Condition} & \multicolumn{2}{c}{Zero energy} & \multicolumn{2}{c}{Not zero-energy} \\
 & \multicolumn{1}{c}{time{[}s{]}} & \multicolumn{1}{c}{rate} & \multicolumn{1}{c}{time{[}s{]}} & \multicolumn{1}{c}{rate} \\ \hline \hline
Baseline method & 1500 & 4.63\% & 30899 & 95.37\% \\ \hline
Proposed method (Aug 28th) & 32021 & 98.88\% & 380 & 1.17\% \\
Proposed method (Aug 30th) & 32354 & 99.85\% & 47 & 0.15\% \\
Proposed method (Sep 1st) & 32342 & 99.77\% & 74 & 0.23\% \\
Proposed method (Sep 6th) & 32261 & 99.82\% & 59 & 0.18\% \\ \hline
Proposed method (Avg) & 32261 & 99.57\% & 140 & 0.43\% \\ \hline
\end{tabular}
 \vspace{-5mm}
\label{tab:activation_ratio}
\end{table}

\section{Conclusion\add{s} and future work} \label{sec:conclusion}
In this paper, we proposed ZEL, which is a net-zero-energy lifelogging system for office workers that employs heterogeneous harvesters and uses their harvested energy to record lifelogging information at zero-energy consumption levels.
\erase{
Three activity types are recorded: when, where, and what activity.
The ZEL system also incorporates a dual power switching circuit and comparators for system state scheduling to minimize battery usage.
In the lifelog generation block, the recorded time-series data are used to generate a user's lifelog by recognizing locations and activities via ML.
As part of our investigations, 
We implemented the ZEL system on a nametag-shaped wearable device and conducted a large-scale data collection experiment with 11 participants.
}
\add{
To evaluate the ZEL, we conducted a large-scale data collection experiment with 11 participants.
}
The obtained results show we achieved an 87.2\% accuracy level for 8-place recognition and 93.1\% accuracy level for static/dynamic recognition using the PD model.
\add{
We also showed that heterogeneous harvesters improve recognition accuracy.
}
In addition, we demonstrated that ZEL could function as a net-zero-energy system by achieving an approximate 99\% zero-energy rate\add{s} in a real environment through the use of a battery minimization mechanism.

\erase{
The ZEL system has a few challenges to overcome before it can be considered user-friendly.
The first issue is the size and weight of the device.
Our nametag-shaped device is 123 mm square, which is too large for a wearable, and weighs 192 grams.
This may be too heavy for some users and might cause neck pain.
However, the circuit is small enough, and the size and weight can be reduced by using a smaller and more lightweight harvester\cite{doi:10.1021/acsami.9b00018}. 

The second issue is versatility.
As shown in Section~\ref{sec:investigation}, the recognition accuracy of the PI model is significantly reduced.
This could be caused by shadows created by the user's posture or by the orientation of the light source.
These factors can be overcome by changing the device placement.
For example, if the system is implemented in a hat-shaped device, recognition accuracy could be improved because it would not be affected by the user's posture or orientation.

Finally, there is the issue of data acquisition.
In this ZEL system prototype, users need to connect a cable to the device and extract the data when they want to check their lifelog.
In the future, we will simplify this data acquisition process by incorporating a program into the onboard device that performs context recognition processing and then wirelessly sends the recognition results to a smartphone.
The current ZEL prototype is designed to consume higher energy than required for the end-to-end context recognition shown in SolAR\cite{Sandhu2021SolAREP} by intentionally lighting up the LED.
As a result, we can conclude that the ZEL prototype has sufficient surplus energy to implement end-to-end context awareness recording and lifelog information transmission.
}
\add{
The ZEL system has some challenges.
Especially, there is the issue of data acquisition.
In this ZEL system prototype, users need to connect a cable to the device and extract the data.
In the future, we will simplify this data acquisition process by wireless communication.
}



 \section*{Acknowledgment}
 This work was partially supported by the Japan Society for the Promotion of Science (JSPS) KAKENHI Grant Numbers JP18H03233 and JP19H05665.

\bibliographystyle{IEEEtran}

\bibliography{ref}

\end{document}